# On the Sum of Squared $\eta$-$\mu$ Random Variates With Application to the Performance of Wireless Communication Systems

Imran Shafique Ansari, *Student Member, IEEE*, Ferkan Yilmaz, *Member, IEEE*, and Mohamed-Slim Alouini, *Fellow, IEEE*

*Abstract*—The probability density function (PDF) and cumulative distribution function of the sum of $L$ independent but not necessarily identically distributed squared $\eta$-$\mu$ variates, applicable to the output statistics of maximal ratio combining (MRC) receiver operating over $\eta$-$\mu$ fading channels that includes the Hoyt and the Nakagami-$m$ models as special cases, is presented in closed-form in terms of the Fox's $\bar{H}$ function. Further analysis, particularly on the bit error rate via PDF-based approach, is also represented in closed form in terms of the extended Fox's $\bar{H}$ function ($\hat{H}$). The proposed new analytical results complement previous results and are illustrated by extensive numerical and Monte Carlo simulation results.

## I. INTRODUCTION

IN recent times, different diversity schemes have marked an important impact in the arena of wireless communication systems. The main reason behind this is that these different diversity schemes allow for multiple transmission and/or reception paths for the same signal [1]. The optimal diversity combining scheme is the maximal ratio combining (MRC) diversity scheme where all the diversity branches are processed to obtain the best possible signal-to-noise ratio (SNR) [1]–[3]. This results into extensive occurrence of the statistical distribution of the sum of squared envelopes of faded signals in several wireless communication systems [4].

Additionally, wireless communications are driven by a complicated phenomenon known as radio-wave propagation that is characterized by various effects such as fading, shadowing and path-loss. The statistical behavior of these effects is described by different models depending on the nature of the communication environment. Various distributions have several applications in wireless communication engineering problems and one of those that we focus on is a communication system employing MRC diversity scheme undergoing $\eta$-$\mu$[1] distribution i.e. the study of MRC diversity combining

Imran Shafique Ansari, Ferkan Yilmaz and Mohamed-Slim Alouini are with the Computer, Electrical, and Mathematical Sciences and Engineering (CEMSE) Division at King Abdullah University of Science and Technology (KAUST), Al-Khawarizmi Applied Math. Building (Bldg. #1), Thuwal 23955-6900, Makkah Province, Kingdom of Saudi Arabia (e-mail: {imran.ansari, ferkan.yilmaz, slim.alouini}@kaust.edu.sa).

[1]In this work, both the Formats are considered and hence the results presented herein are applicable to both the Formats.

receiver operating over $\eta$-$\mu$ fading channels [5], where the statistics of the sum of squared $\eta$-$\mu$ random variates (RVs) are required. Moreover, the performance analysis of such wireless communication systems usually requires complicated and tedious tasks related to statistics as explained in detail in [6]. For instance, it is worth sharing here that the probability density function (PDF) and cumulative distribution function (CDF) of the sum of $L$ independent but not necessarily identical (i.n.i.d.) Gamma RVs have been investigated in [7] using the results presented in [8] to address the performance of MRC diversity receivers over Nakagami-$m$ fading channels. Similarly, in [9]–[11], the authors have independently presented a closed-form expression for the PDF of the sum of independent Gamma variates with arbitrary fading parameters.

Recently, the $\eta$-$\mu$ fading distribution that includes the Nakagami-$m$ and the Hoyt distribution as its special cases, has been proposed as a more flexible model for practical fading radio channels [12] and free-space optical (FSO) communication systems. Additionally, the $\eta$-$\mu$ model has been found suitable for modeling small scale fading phenomenon in non-line-of-sight mobile communication channels [12]. Non-homogeneous physical characteristics of the environment have been taken into consideration in the mathematical modeling of this channel that sometimes best fits to the practical scenario and experimental data [13]. The $\eta$-$\mu$ distribution fits well to experimental data and can accurately approximate the sum of i.n.i.d. Hoyt envelopes having arbitrary fading degrees [14]. In particular, its tail closely follows the true statistics where other distributions fail to yield a good fit [15]. Beyond that, this distribution has been gaining interest in the field of performance evaluation of digital communications over fading channels (for example, see [4] and references therein). Furthermore, in [4], the authors have published results on the distribution of the sum of the squared envelopes of $\eta$-$\mu$ faded signals but the published results are given in terms of nested infinite-series summations representation, dependent on the number of channel branches and utilizing the confluent Lauricella hypergeometric function [16]. It has been stated in [17] and other relevant research proceedings that the confluent form of the multivariate Lauricella hypergeometric function is inherently difficult to be evaluated especially when its order increases. Moreover, this function does not have a simple real-axis integral representation.

In this work, we offer novel closed-form expressions for the PDF and CDF of the sum of i.n.i.d. squared $\eta$-$\mu$ RVs in terms of easily computable Fox's H̄ function [18]–[20, App. (A.5)]. It is noteworthy to mention that the bit error rate (BER) is one of the most important performance measures that forms the basis in designing wireless communication systems. Hence, we also propose new closed-form expressions of the BER, as a performance metric, for binary modulation schemes, via a PDF-based approach, of a *L*-branch MRC diversity receiver in the presence of $\eta$-$\mu$ multipath fading, in terms of the extended Fox's H̄ function ($\hat{H}$) [21][2]. This proves the importance and the simplicity in the employment of those earlier derived simple closed-form statistical PDF and CDF expressions. These resulting easy-to-evaluate expressions also give an alternative form for previously known/published results. It should be noted that all our newly proposed results are readily computable using the Mellin-Barnes theorem that further corroborates the generality and the usefulness of the analytical frameworks introduced in this paper. These have been checked and validated by Monte Carlo simulations.

The remainder of the paper is organized as follows. Section II introduces the system and Section III gives novel closed-form expressions for the PDF and CDF of the sum of squared $\eta$-$\mu$ RVs in terms of the Fox's H̄ function. Next, Section IV utilizes these results presented in Section III to derive useful expressions for the BER, as a performance metric for MRC diversity receivers operating over i.n.i.d. $\eta$-$\mu$ fading channels or diversity paths in terms of the extended Fox's H̄ function ($\hat{H}$). Further, Section IV also discusses the results followed by the summary of the paper in the last section.

## II. CHANNEL AND SYSTEM MODELS

A MRC based communication system with a source and a destination is considered with *L* diversity paths undergoing i.n.i.d. $\eta$-$\mu$ fading channels as follows

$$Y_l = \alpha_l X + n_l, \quad l = 1, 2, \ldots, L, \quad (1)$$

where $Y_l$ is the received signal at the receiver end, $X$ is the transmitted signal, $\alpha_l$ is the channel gain, and $n_l$ is the additive white Gaussian noise (AWGN). In an $\eta$-$\mu$ multipath fading channel, $\gamma_l = |\alpha_l|^2$ follows squared $\eta$-$\mu$ distribution. Hence, the channel gains experience multipath fading whose statistics follow squared $\eta$-$\mu$ distribution with PDF given by

$$p_{\gamma_l}(\gamma) = \frac{2\sqrt{\pi}\mu_l^{\mu_l+\frac{1}{2}} h_l^{\mu_l}}{\Gamma(\mu_l) H_l^{\mu_l-\frac{1}{2}}} \frac{\gamma_l^{\mu_l-\frac{1}{2}}}{\bar{\gamma}_l^{\mu_l+\frac{1}{2}}} \exp\left(-\frac{2\mu_l h_l \gamma_l}{\bar{\gamma}_l}\right)$$
$$\times I_{\mu_l-\frac{1}{2}}\left(\frac{2\mu_l H_l \gamma_l}{\bar{\gamma}_l}\right), \quad (2)$$

where $\mu_l > 0$ is known as fading figure representing the diversity order of the fading environment, $\Gamma(\cdot)$ denotes the Gamma function [22, Eq. (8.310)], $I_v(.)$ is the modified Bessel function of the first kind and arbitrary order $v$ [22, Eq. (8.406.1)], and $\bar{\gamma}_l = \mathbb{E}[\gamma_l]$ is the average SNR with $\mathbb{E}[.]$ denoting expectation. In $\eta$-$\mu$ fading model, $2\mu_l$ represents the number of multipath clusters [12], [13].

The PDF of $\gamma_l$ can be expressed in two formats, described in what follows.

• *The $\eta$-$\mu$ Fading Model - Format 1*: In Format 1, the parameter $0 < \eta_l < \infty$ denotes the scattered-wave power ratio in a non-homogeneous environment between the in-phase and quadrature components of the fading signal within each cluster of multipath that are assumed to be independent from each other and to have different powers. In such a case, $h_l = \frac{2+\eta_l^{-1}+\eta_l}{4}$ and $H_l = \frac{\eta_l^{-1}-\eta_l}{4}$. It is noted that in Format 1, $H/h = (1-\eta)/(1+\eta)$.

• *The $\eta$-$\mu$ Fading Model - Format 2*: In Format 2, $-1 < \eta_l < 1$ denotes the correlation coefficient in a non-homogeneous environment between the scattered-wave powers of the in-phase and quadrature components of each multipath cluster. In such a case, $h_l = \frac{1}{1-\eta_l^2}$ and $H_l = \frac{\eta_l}{1-\eta_l^2}$. It is noted that in Format 2, $H/h = \eta$.

It can be easily seen from above that one Format may be obtained from other by the bilinear relation $\eta_1 = (1-\eta_2)/(1+\eta_2)$ and/or equivalently $\eta_2 = (1-\eta_1)/(1+\eta_1)$, where $\eta_1$ is the parameter $\eta$ for Format 1 and $\eta_2$ is the parameter $\eta$ for Format 2 [15].

As discussed earlier, the $\eta$-$\mu$ fading distribution comprises both Hoyt ($\mu_l = 0.5$) and Nakagami-$m$ ($\eta_l \to 0, \eta_l \to \infty, \eta_l \to \pm 1$) as its special cases.

In a MRC combiner, received signals from all diversity antennas are algebraically added together and the combiner is connected to a suitable detector to detect the information signal. The output SNR of the combiner can be expressed by [1]

$$\gamma_{MRC} = \sum_{l=1}^{L} \gamma_l = \frac{E_b}{N_0} \sum_{l=1}^{L} |\alpha_l|^2, \quad (3)$$

where the instantaneous SNR of the *l*th branch is given by $\gamma_l = (E_b/N_0)|\alpha_l|^2$, $E_b$ is the average energy per bit, and $N_0$ is the one sided power spectral density of the AWGN.

## III. CLOSED-FORM STATISTICAL CHARACTERISTICS FOR THE SUM OF SQUARED $\eta$-$\mu$ RANDOM VARIATES

This section presents the results on the statistical characteristics including the PDF and CDF of the sum of i.n.i.d. squared $\eta$-$\mu$ variates.

**Theorem 1** (PDF of the sum of squared $\eta$-$\mu$ RVs). *Let $\{\gamma_l\}_{l=1}^{L}$ be a set of i.n.i.d. squared $\eta$-$\mu$ variates with parameters $\mu_l$ and $\eta_l$. Then, the closed-form PDF of the sum*

$$Y = \sum_{l=1}^{L} \gamma_l \quad (4)$$

*can be expressed in terms of Fox's H̄ function[3] as*

$$p_Y(y) = \bar{H}_{2L,2L}^{0,2L}\left[\exp(y) \,\middle|\, \begin{matrix} \Xi_L^{(1)}, \Xi_L^{(2)} \\ \Xi_L^{(3)}, \Xi_L^{(4)} \end{matrix}\right], \quad (5)$$

*where $y > 0$, the coefficient sets $\Xi_k^{(1)}$, $\Xi_k^{(2)}$, $\Xi_k^{(3)}$, and $\Xi_k^{(4)}$, $k \in \mathbb{N}$ are defined as*

$$\Xi_k^{(1)} = (0, A_1, \mu_1), \ldots, (0, A_k, \mu_k), \quad (6)$$

---

[2]The extended Fox's H̄ function ($\hat{H}$), the Fox's H̄ function, the Fox's H function, and the Meijer's G function are extensively defined in [9], [10, Table I].

[3]An efficient MATHEMATICA® implementation of the Fox's H̄ function is available in [9], [10, Table II] (similar to [23]–[25]).

$$\Xi_k^{(2)} = (0, B_1, \mu_1), \ldots, (0, B_k, \mu_k), \quad (7)$$

$$\Xi_k^{(3)} = (-1, A_1, \mu_1), \ldots, (-1, A_k, \mu_k), \quad (8)$$

and

$$\Xi_k^{(4)} = (-1, B_1, \mu_1), \ldots, (-1, B_k, \mu_k), \quad (9)$$

respectively with $A_l = \frac{\bar{\gamma}_l}{2\mu_l(h_l - H_l)}$ and $B_l = \frac{\bar{\gamma}_l}{2\mu_l(h_l + H_l)}$.

*Proof:* In order to derive the PDF of $Y$, we proceed as follows. Firstly, the MGF

$$\mathcal{M}_{\gamma_l}(s) \triangleq \mathbb{E}\left[\exp(-\gamma_l s)\right] = \int_0^\infty \exp(-\gamma_l s) p_{\gamma_l}(\gamma) d\gamma \quad (10)$$

of a single $\eta$-$\mu$ distribution is given as [4]

$$\mathcal{M}_{\gamma_l}(s) = (1 + sA_l)^{-\mu_l}(1 + sB_l)^{-\mu_l}. \quad (11)$$

Then, after performing some simple algebraic manipulations using [26, Eq. (6.1.15)], the MGF of a single $\eta$-$\mu$ distribution can be rewritten as

$$\mathcal{M}_{\gamma_l}(s) = \frac{\Gamma^{\mu_l}(1 + sA_l)}{\Gamma^{\mu_l}(2 + sA_l)} \frac{\Gamma^{\mu_l}(1 + sB_l)}{\Gamma^{\mu_l}(2 + sB_l)}. \quad (12)$$

Since, $\gamma_l's$ are independent, the MGF of Y is the product of the MGF's of the $\gamma_l's$, i.e.

$$\mathcal{M}_Y(s) = \prod_{l=1}^L \mathcal{M}_{\gamma_l}(s). \quad (13)$$

Now, the PDF of the sum of squared $\eta$-$\mu$ RVs, using the obtained MGF in (13), via inverse Laplace transform [27] can be expressed as

$$p_Y(y) = \mathcal{L}^{-1}\{\mathcal{M}(s)\} = \frac{1}{2\pi i} \oint_C \mathcal{M}_Y(s) \exp(s) ds \quad (14)$$

that produces a Mellin-Barnes contour integral [28] representation as

$$p_Y(y) = \frac{1}{2\pi i} \oint_C \prod_{l=1}^L \frac{\Gamma^{\mu_l}(1 + sA_l)}{\Gamma^{\mu_l}(2 + sA_l)} \frac{\Gamma^{\mu_l}(1 + sB_l)}{\Gamma^{\mu_l}(2 + sB_l)} \exp(s) ds. \quad (15)$$

Hence we use this obtained result and perform some simple rearrangements on the $\Gamma(.)$ terms in the Mellin-Barnes contour integral representation to express the closed-form PDF of the sum of squared $\eta$-$\mu$ RVs, $Y$, in terms of the Fox's $\bar{\text{H}}$ function as given in (5). ∎

It is worthy mentioning that the PDF of sum of squared $\eta$-$\mu$ RVs has also been successfully achieved in terms of the confluent form of the multivariate Lauricella hypergeometric function in [4] that typically involves an $L$-fold infinite summations. On the other hand, our alternative result presented in Theorem 1 involves only one single-fold integration.

The motivation and possibility that led to the above result presented in Theorem 1 was the representation of the PDF of a single squared $\eta$-$\mu$ RV in terms of the Fox's $\bar{\text{H}}$ function as discussed below in Corollary 1.

**Corollary 1** (PDF of a single squared $\eta$-$\mu$ RV). *Let $\{\gamma_l\}$ be a squared $\eta$-$\mu$ variate with parameters $\mu_l$ and $\eta_l$. Then,* the closed-form PDF of this single squared $\eta$-$\mu$ RV can be expressed as

$$p_{\gamma_l}(\gamma_l) = \bar{\text{H}}_{2,2}^{0,2}\left[\exp(\gamma_l) \middle| \begin{array}{c} (0, A_l, \mu_l), (0, B_l, \mu_l) \\ (-1, A_l, \mu_l), (-1, B_l, \mu_l) \end{array}\right], \quad (16)$$

where $A_l$ and $B_l$ are as given under Theorem 1.

*Proof:* The PDF of a single squared $\eta$-$\mu$ RV can be expressed using the obtained MGF in (12), via inverse Laplace transform [27] that produces a Mellin-Barnes integral [28] representation. Hence we use this obtained result to express the PDF of a single squared $\eta$-$\mu$ RV in an alternative form, in terms of the Fox's $\bar{\text{H}}$ function, as expressed in (16). ∎

Fig. 1 presents the PDF of the output SNR obtained from the exact closed-form expression (5) and shows a perfect match between this obtained closed-form analytical result and the one obtained via Monte Carlo simulations for varying $L's$ (i.e. $L = 3, 4, 5$), and their respective fixed fading parameters $\mu_1 = 1$, $\mu_2 = 1.5$, $\mu_3 = 2$, $\mu_4 = 3.5$, and $\mu_5 = 4.5$. The value of $\eta$ was fixed at $1.2$ representing Format 1. Similar results have also been obtained for Format 2.

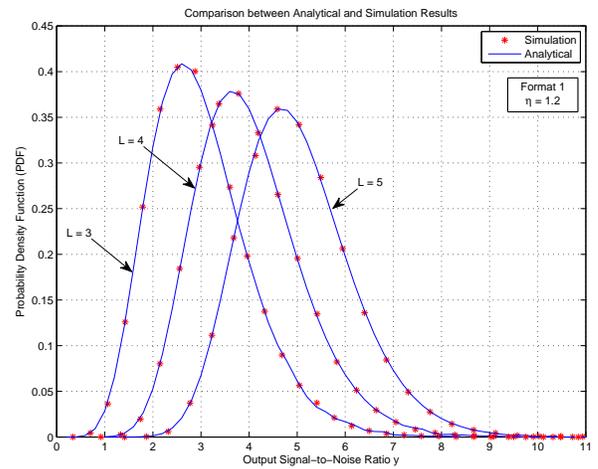

Fig. 1. Comparison between PDFs obtained analytically and via Monte Carlo simulations for varying branches $L$ and respective fixed fading parameters for these channels as $\mu_1 = 1$, $\mu_2 = 1.5$, $\mu_3 = 2$, $\mu_4 = 3.5$, and $\mu_5 = 4.5$.

**Theorem 2** (CDF of the sum of squared $\eta$-$\mu$ RVs). *The CDF of $Y$ can be closely expressed in terms of the Fox's $\bar{\text{H}}$ function as*

$$P_Y(y) = 1 + \bar{\text{H}}_{2L+1,2L+1}^{0,2L+1}\left[\exp(y) \middle| \begin{array}{c} \Xi_L^{(1)}, \Xi_L^{(2)}, (1,1,1) \\ \Xi_L^{(3)}, \Xi_L^{(4)}, (0,1,1) \end{array}\right], \quad (17)$$

where the coefficient sets $\Xi_k^{(1)}$, $\Xi_k^{(2)}$, $\Xi_k^{(3)}$ and $\Xi_k^{(4)}$ are defined earlier in (6), (7), (8), and (9), respectively.

*Proof:* In order to derive the CDF of $Y$, we proceed as follows.

We integrate the PDF expressed in (16) from 0 through $\gamma$ and obtain the CDF for a single squared $\eta$-$\mu$ RV, in terms of the Fox's $\bar{H}$ function, as

$$P_{\gamma_l}(\gamma_l) = \bar{H}_{3,3}^{0,3}\left[\exp(\gamma_l) \left| \begin{array}{c} (0, A_l, \mu_l), (0, B_l, \mu_l), (1,1,1) \\ (-1, A_l, \mu_l), (-1, B_l, \mu_l), (0,1,1) \end{array}\right.\right], \quad (18)$$

where $A_l$ and $B_l$ are as given under Theorem 1.

Now, performing a similar integral operation on (5), utilizing a similar explanation as presented in the proof of the PDF of the sum of squared $\eta$-$\mu$ RVs i.e. Theorem 1 to obtain (5) from (13), and further making some simple modifications to the Mellin-Barnes integral representation to satisfy the exact definition of the Fox's $\bar{H}$ function, we obtain a final closed-form result for the CDF of $Y$, in terms of Fox's $\bar{H}$ function as presented in (17). Hence, in other words, the expression presented in (18) is a special case of (17) in Theorem 2 for $L = 1$. ∎

Fig. 2 and Fig. 3 present the CDF and the logarithmic CDF respectively of the output SNR obtained from the exact closed-form expression (17) and show a perfect match between this obtained closed-form analytical result and the one obtained via Monte Carlo simulations for varying $L's$ (i.e. $L = 2, 3, 4, 5$), and their respective fixed fading parameters $\mu_1 = 1$, $\mu_2 = 1.5$, $\mu_3 = 2$, $\mu_4 = 3.5$, and $\mu_5 = 4.5$. The value of $\eta$ was fixed at $1.2$ representing Format 1. Similar results have also been obtained for Format 2.

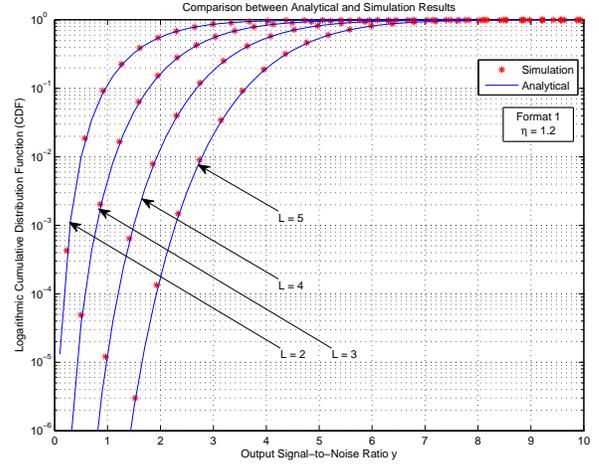

Fig. 3. Comparison between CDFs obtained analytically and via Monte Carlo simulations, on log scale, for varying branches $L$ and respective fixed fading parameters for these channels as $\mu_1 = 1$, $\mu_2 = 1.5$, $\mu_3 = 2$, $\mu_4 = 3.5$, and $\mu_5 = 4.5$.

### A. Outage Probability

When the probability that the instantaneous MRC output SNR falls below a given threshold $y_{\text{th}}$, we encounter a situation labeled as outage and it is an important feature to study OP of a system. Hence, another important fact worth stating here is that the expression derived in the Theorem 2 also serves the purpose for the expression of OP of MRC diversity combining receivers based wireless communication system that is experiencing i.n.i.d. $\eta$-$\mu$ fading channels or in other words, when the desired user is subject to $\eta$-$\mu$ fading, the probability that the SNR falls below a predetermined protection ratio $y_{\text{th}}$ can be simply expressed by replacing $y$ with $y_{\text{th}}$ in (17) as

$$P_{\text{out}}(y_{\text{th}}) = P_Y(y_{\text{th}}). \quad (19)$$

Employing similar substitutions, all the other respective expressions of CDF can be utilized for OP such as replacing $\gamma_l$ with $\gamma_{\text{th}}$ in (18) and/or replacing $y$ with $y_{\text{th}}$ in (17).

### B. Average BER

The most straightforward approach to obtain BER $P_e$ for MRC is to average the conditional BER [4] for the given SNR, over the PDF of the combiner output SNR i.e. $P_e = \int_0^\infty P_e(\epsilon|y) p_Y(y) dy$ [1].

**Theorem 3** (BER of a $L$-branch MRC system operating over $\eta$-$\mu$ fading channels for binary modulation schemes). *The BER of a $L$-branch MRC diversity combining receiver wireless communication system running over $\eta$-$\mu$ fading channels, valid for any binary modulation scheme including coherent*

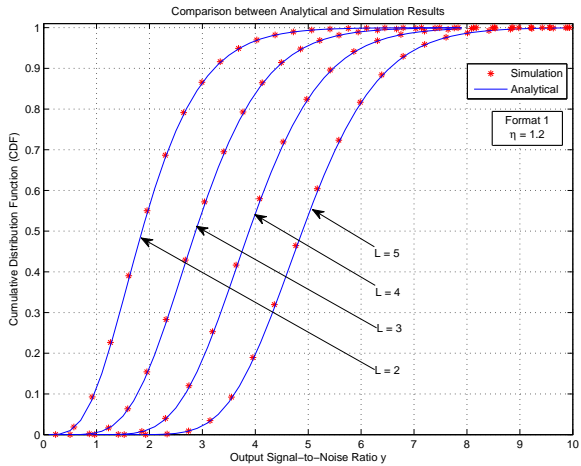

Fig. 2. Comparison between CDFs obtained analytically and via Monte Carlo simulations for varying branches $L$ and respective fixed fading parameters for these channels as $\mu_1 = 1$, $\mu_2 = 1.5$, $\mu_3 = 2$, $\mu_4 = 3.5$, and $\mu_5 = 4.5$.

## IV. APPLICATIONS TO THE PERFORMANCE OF DIVERSITY COMBINING RECEIVER SYSTEMS

This section presents the results on the performance analysis, in particular outage probability (OP) and BER analysis, of the sum of i.n.i.d. squared $\eta$-$\mu$ variates. In here, we limit ourselves to binary modulation schemes.

In MRC combining scheme, all the branches are selected at the output. In our case, for a $L$-branch MRC diversity receiver, the SNR $y$, is given by (3).

---

[4]The expression $P_e(\epsilon|y) = \frac{\Gamma(p, qy)}{2\Gamma(p)}$ is a unified conditional BER expression for coherent and non-coherent binary modulation schemes over an AWGN channel [29]. $\Gamma(\cdot, \cdot)$ is the complementary incomplete gamma function [22, Eq. (8.350.2)]. The parameters $p$ and $q$ account for different modulation schemes. For an extensive list of modulation schemes represented by these parameters, one may look into [30] or refer to Table I.

TABLE I
CONDITIONAL BIT ERROR PROBABILITY (BER) PARAMETERS

| Modulation | $p$ | $q$ |
|---|---|---|
| Coherent Binary Frequency Shift Keying (CBFSK) | 0.5 | 0.5 |
| Coherent Binary Phase Shift Keying (CBPSK) | 0.5 | 1 |
| Non-Coherent Binary Frequency Shift Keying (NBFSK) | 1 | 0.5 |
| Differential Binary Phase Shift Keying (DBPSK) | 1 | 1 |

*binary frequency shift keying (CBFSK), non-coherent binary frequency shift keying (NBFSK), coherent binary phase shift keying (CBPSK), and differential binary phase shift keying (DBPSK), can be expressed in closed-form, in terms of the extended Fox's $\bar{H}$ function ($\hat{H}$)[5], as*

$$P_e = \frac{q^p}{2}\hat{H}^{2L+1,1}_{2L+2,2L+2}\left[1 \left| \begin{array}{c}(1-q,1,p),\zeta_1 \\ \zeta_2,(-q,1,p)\end{array}\right.\right], \quad (20)$$

*where*

$$\zeta_1 = \Upsilon^{(1)}_L, \Upsilon^{(2)}_L, (1,1,1), \quad (21)$$

*and*

$$\zeta_2 = (0,1,1), \Upsilon^{(3)}_L, \Upsilon^{(4)}_L, \quad (22)$$

*and where the coefficient sets $\Upsilon^{(1)}_k$, $\Upsilon^{(2)}_k$, $\Upsilon^{(3)}_k$ and $\Upsilon^{(4)}_k$, $k \in \mathbb{N}$ are defined as*

$$\Upsilon^{(1)}_k = (2,A_1,\mu_1),\ldots,(2,A_k,\mu_k), \quad (23)$$

$$\Upsilon^{(2)}_k = (2,B_1,\mu_1),\ldots,(2,B_k,\mu_k), \quad (24)$$

$$\Upsilon^{(3)}_k = (1,A_1,\mu_1),\ldots,(1,A_k,\mu_k), \quad (25)$$

*and*

$$\Upsilon^{(4)}_k = (1,B_1,\mu_1),\ldots,(1,B_k,\mu_k), \quad (26)$$

*respectively with $A_l$ and $B_l$ are as defined in Theorem 1.*

*Proof:* Utilizing BER equation $P_e$ by substituting the conditional BER equation $P_e(\epsilon|y)$ and (5) into it and performing some simple manipulations along with some simple rearrangements of $\Gamma(.)$ function terms, we get an exact closed-form result of the integral valid for any binary modulation scheme including CBFSK, NBFSK, CBPSK, and DBPSK, in terms of the extended Fox's $\bar{H}$ function ($\hat{H}$), as presented above in (20), Theorem 3. ∎

### C. Results and Discussion

The numerical results for BER of MRC diversity combining receiver scheme with $L$-diversity over i.n.i.d. $\eta$-$\mu$ fading channels are presented in this section.

The average SNR per bit in all the scenarios discussed is assumed to be equal. In addition, different digital modulation schemes are represented based on the values of $p$ and $q$ where $p = 0.5$ and $q = 1$ represents CBPSK, $p = 1$ and $q = 1$ represents DBPSK, CBFSK is represented by $p = 0.5$ and $q = 0.5$, and NBFSK is represented by $p = 1$ and

[5]The extended Fox's $\bar{H}$ function ($\hat{H}$) was first introduced in [21] and has a MATHEMATICA® implementation given in [9], [10, Table III].

$q = 0.5$. In Monte Carlo simulations, the $\eta$-$\mu$ fading channel was generated based on the description given in [12].

We observe from Fig. 4 that the analytical results provide a perfect match to the MATLAB simulated results and the results are as expected i.e. the BER decreases as the SNR increases. Its important to note here that these values for the parameters were selected randomly to prove the validity of the obtained results and hence specific values based on the standards can be used to obtain the required results.

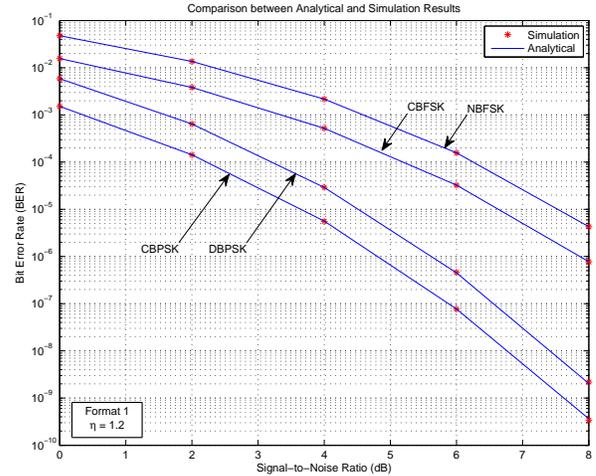

Fig. 4. Average BER of different binary modulation schemes over i.n.i.d. $\eta$-$\mu$ fading channels with $L = 5$-branch MRC and fading parameters for these channels as $\mu_1 = 1$, $\mu_2 = 1.5$, $\mu_3 = 2$, $\mu_4 = 3.5$, and $\mu_5 = 4.5$.

Furthermore, it can be seen from Fig. 4 that, as expected, CBPSK outperforms the other modulation schemes and the coherent binary modulation schemes outperform their respective non-coherent and/or differential binary modulation scheme i.e. CBPSK outperforms DBPSK and CBFSK outperforms NBFSK. Additionally, PSK in general performs better than FSK, as expected. Similar results for any other values of $\mu$'s and $\eta$'s can be observed for the exact closed-form BER for $L$-diversity i.n.i.d. $\eta$-$\mu$ channels presented in this work.

## V. CONCLUDING REMARKS

We derived novel closed-form expressions for the PDF and the CDF of the sum of i.n.i.d. squared $\eta$-$\mu$ RVs. An interesting finding is that these expressions can be written in terms of special functions, specifically the Fox's $\bar{H}$ function distributions. Based on these statistical formulas obtained, and following the exact PDF-based approach, we analyzed the performance of a MRC diversity combining receiver based wireless communication system operating over i.n.i.d. $\eta$-$\mu$ fading channels and important performance metrics such as OP and BER were expressed in closed form and hence this serves as the key feature along side the novel statistical derivations of PDF and CDF.

An exact closed-form expression for the BER performance of different binary modulations with $L$-branch MRC scheme over i.n.i.d. $\eta$-$\mu$ fading channels was derived. The analytical calculations were done utilizing a general class of special functions including the Fox's H function [31], the Fox's H̄ function, and the extended Fox's H̄ function ($\hat{H}$). In addition, this work presents numerical examples to illustrate the mathematical formulations developed in this work and to show the effect of the fading severity and unbalance on the system performance. Our results complement previously published results that are in the form of infinite sums or nested sums of the fading parameter.

### ACKNOWLEDGMENT

We would like to thank King Abdullah University of Science and Technology (KAUST), Thuwal, Makkah Province, Saudi Arabia for providing support and resources for this research work.